# QUANTIZATION OF MASSES IN THE SOLAR SYSTEM.

By A. M. Ilyanok and I. A. Timoshchenko.

*The hypothesis of quantization of masses of the solar system planets is considered. It is supposed that the solar system was bearing during the Protosun squeezing from the red giant to a yellow dwarf. According to mass and orbit redistribution of interior planets the condition of the origin of satellites of internal and external planets located in the ecliptic plane are found. It is shown, that the main source of substance for these satellites and majority of meteorites and asteroids in the solar system is mantle and bark of the Mars separated from it under influence of gravitation forces about 4 billion years ago, caused by squeezing of the sun and Jupiter. In these condition there is a probability of contamination of the Earth by DNA of living organisms from Mars. It is shown that Uranus and Neptune were formed as a result of protosaturn decay. It is found that from all external planets only Uranus has the large cavity inside. As a result there is no internal power source, and the considerable shift of a magnetic axis relative to rotate axis is observed.*

**1. Introduction.**

In paper [1] the new approach to describing physical phenomena is given. For comparison of experimental data in different branches of physics the universal measuring tool - dimensionless scale is used. It is following:

$$L_\alpha = b \frac{n_1}{n_2} \alpha^n + b^* \frac{n_1^*}{n_2^*} \alpha^{n^*}, \tag{1}$$

where $n = \pm 0, 1, 2, 3,...$; $n_1 = 0, 1, 2, 3,...$; $n_2 = 0, 1, 2, 3,...$;

$b = 0, 1, (2\pi)^{-1/2}, (\pi)^{-1/2}, (2\pi)^{-1}$;

$n^* = \pm 0, 1, 2, 3,...$; $n_1^* = \pm 0, 1, 2, 3,...$; $n_2^* = 0, 1, 2, 3,...$;

$b^* = 0, 1, (2\pi)^{-1/2}, (\pi)^{-1/2}, (2\pi)^{-1}$;

$\alpha^{-1} = 137.0360547255...$ is fine structure constant.

The scale is transformed to scale with necessary dimension by multiplication $L_\alpha$ on elementary charge $e$, Planck constant $h$, light velocity $c$, etc.

On the base of scale (1) scale of planet's orbit radii and their average orbital velocities were calculated:

$$R_n = \left(\frac{n + 2(2m+1)}{3}\right)^2 R_1 \tag{2}$$

$$v_n = \frac{3v_1}{n + 2(2m+1)}, \tag{3}$$

where $n = 1, 2, 3, 4, 5, 6, 7, 8, 9$, $m = 0, 0, 0, 0, 1, 2, 3, 4, 5$; $v_1$ and $R_1$ are average orbital velocity and big semiaxis of the Mercury's orbit. It is connected with fundamental constant by the following way:

$$v_1 = 3\alpha^2 c = 47.89307 \text{ km/sec}. \tag{4}$$

$$R_1 = \frac{h}{\alpha^{12} m_p c} = 5.796 \cdot 10^{10} \text{ m}. \tag{5}$$

where $m_p$ is proton mass.

Let us calculate planet's orbit radii and its average orbital velocities using formulas (2) and (3). Compare our calculation with experimental data. This data are represented in the Table.

Table. Calculation of average velocities and major semiaxis of planets in the solar system.

| № | Planet | Experimental average orbital velocity $v_n^*$, km/sec | Theoretical average orbital velocity $v_n^*$, km/sec | Difference between experimental and theoretical value $\delta$, % | Experimental value of big semiaxis of orbit $R_n^*$, ($\times 10^6$ km) | Theoretical value of big semiaxis of orbit $R_n^*$, ($\times 10^6$ km) | Difference between experimental and theoretical value $\delta$, % |
|---|--------|------|------|--------|--------|--------|--------|
| 1 | Mercury | 47.89 | 47.893 | +0.0064 | 57.90 | 57.96 | + 0.10 |
| 2 | Venus | 35.03 | 35.919 | +2.50 | 108.20 | 103.04 | - 5.0 |
| 3 | Earth | 29.79 | 28.74 | -3.60 | 149.6 | 161.0 | +10.76 |
| 4 | Mars | 24.13 | 23.95 | -0.75 | 227.9 | 231.84 | +1.73 |
| 5 | Jupiter | 13.06 | 13.0617 | +0.013 | 778.3 | 779.24 | + 0.120 |
| 6 | Saturn | 9.64 | 8.980 | -7.30 | 1427.0 | 1648.6 | +15.50 |
| 7 | Uranus | 6.81 | 6.841 | +0.46 | 2869.6 | 2840.0 | -1.0 |
| 8 | Neptune | 5.43 | 5.526 | +1.77 | 4496.6 | 4353.4 | -3.30 |
| 9 | Pluto | 4.74 | 4.635 | -2.20 | 5900.0 | 6188.8 | +4.90 |

**2. Interior planets of the solar system.**

Let us consider, that after formation the planets had velocities and big semiaxis, exactly describing by formulas (2) and (3). According to relations (4) and (5), their values should not vary, because the fundamental constants do not vary. But from present experiment we have displacement of planets and change of their velocities relative to the theoretical values. How it could take place?

The present models of the solar system formation consider its formation from gaseous and dust clouds. Thus, it is considered, that cooling down the sun would turn to the red giant. However there are no direct experimental proof of this model both in the solar system, and on other stars. Therefore alternative versions of the solar system development can have place. One of them was present by Egyed L. Developing his ideas, we suggest that the solar system had originated from center of the Galaxy. Originally it consist of three stars that were red giants − protosun, protojupiter and protosaturn. Protojupiter and protosaturn was separated from protosun and promptly had faded, and transformed to planets consisting mainly of liquid hydrogen. The sun continued to be squeezed, and at present it is a yellow dwarf.

In paper [2] it is shown that the sun represents a hollow star. Its shell has a density of 12.97 g/cm$^3$ and consists of liquid atomic hydrogen with helium dissolved in it. And their gaseous mixture under the pressure of 0.1 atmosphere fills the cavity. At periodic quantum squeezing of protosun it formed new planets in equatorial plane. These planets had an orbital velocity of equal equatorial velocity of protosun at the moment of birth.

It is possible to consider the similar model of satellites formation of external planets. Let us remark, that the planets in this case genetically connect with the sun and rotate in its equator plane. Therefore satellites of external planets, genetically connected with them, should rotate in plane of planet equator. If a satellite is caught from space, it will be rotate in ecliptic plane. At rotational displacement of planet axis, at its motion around the sun satellites only genetically connected with it will be turned round in the equator plane of planet. The entrapped satellites will remain near to a plane of ecliptic. Earlier these facts did not meet a theoretical explanation.

Let us summarize masses of internal planets. It is:

$$M_{Mercury} + M_{Venus} + M_{Earth} + M_{Mars} = 1{,}977\, M_{Earth}$$

Hence, up to a quantum number 2 (which is determined by the scale built in paper [1]) does not suffice

$$0.023\ M_{Earth} = 1.86\ M_{Moon}$$

It is equal to sum of masses of Moon, Triton (satellite of Neptune), Japetus (satellite of Saturn), both meteorites and masses of small planets and meteorites moving in plane of ecliptic. The orbits of a Triton and Japetus are also in ecliptic plane. As such considerable mass could not come from the outside, it is possible to conclude, that it was distributed on internal planets. And after some cataclysm its distribution had become like at present. This mass redistribution had lead to displacement of orbit and average orbital velocities (see Table).

According to the paper [1], at the moment of "birth" the Earth radius was

,

where $G$ is gravitational constant It was less on 115 km the present average radius.

As the mantle density of the Earth (3,4 g/cm$^3$) is close to average density of the Moon (3,34 g/cm$^3$), further all calculations make relative mass and density of the Moon.

The simple calculation has shown that mass of the Earth shell by width of 115 km equal $\Delta M_{Earth} = 2.77 M_{Moon}$. Thus, the Earth, (to fill deficiency in a mass), caught a celestial body. The body with a mass $2.77 M_{Moon}$ was merged by the Earth, and stayed part became the Moon we observe at present.

But whence this mass has appeared? According to fact that the Mercury's orbit was not changed it can be consider that the diameter of the sun was equal to diameter of Mercury's orbit at the moment of a cataclysm, it means that the sun was red giant.

Let us remark that the average density of Mars (3,95 g/cm$^3$) is close to average bark densities of the Earth and the Moon. It can be supposed that the Moon takes its origin from the Mars. What could make the Mars to be torn? The powerful exterior gravitation effect had all chances of it. At present equatorial velocity of the sun rotation less then the Jupiter one in $2\pi$ times. The equatorial velocity of the Jupiter equal $\frac{1}{4}\pi\alpha^2 c = 12.538$ km/sec, and average densities of the sun (1.408 g/cm$^3$) and the Jupiter (1.330 g/cm$^3$) are close. Therefore the Jupiter could be the second star of the solar system. Suppose that protomars located in the mass center of this double star-shaped system, it is possible to find (according to the theory [1]) mass and diameter of the protojupiter at that time. In addition it is supposed that gravitational constant changes in dependence on phase state of substance and its temperature. The ratio of distances from protomars up to protojupiter and from protomars up to the protosun is $\frac{85}{36} = 2.361$. Therefore, the mass of Jupiter before a cataclysm was in 2,361 times less solar one that exceeds a present mass of Jupiter in 443.3 times at the expense of gravitational constant change. If average densities are considered to be invariable, radius of Jupiter in that time exceeded present in $443.3^{1/3} = 7.62$ times, that there is less radius of the nowadays sun on 28 %.

Thus, at the moment of catastrophe in the solar system on the sky of protomars there were two stars: the red giant by the size, equal diameter of orbit of the Mercury, and yellow dwarf with the size, on 28 % smaller present size of the sun. Before cataclysm the protomars had the mass combined from following masses: the present mass of the Mars equal 8.702 $M_{Moon}$, part with mass of 3.776 $M_{Moon}$ which had flown to the Earth and mass of the satellite entrapped in a plane of ecliptic by other planets. This is mass of the Triton equal 0.8 $M_{Moon}$, mass of the Japetus equal 0.026 $M_{Moon}$ and total masses of meteorites and small planets moving in ecliptic plane equal 0.0244 $M_{Moon}$. Masses of natural satellites Mars Phobos and Deymos are neglectfully small. In addition the shape of these satellites is so unsymmetrical, that testifies to origin of them not from liquid matter of mantle, but from a solid bark of protomars. In result it is considered, that the primary mass of protomars was 13.32 $M_{Moon}$. After a disrupture protomars lost a kinetic energy, and its orbit was canted to an equatorial plane of the sun on 5,15° and was displaced to more low level on 3.94 million km. It became promptly rotate around its axis, and the axis turned relatively to orbital plane on 25.

Figure. The Moon motion after its separating from protomars.

In addition the important fact is, that equatorial radius of Mars, equal 3395 km [3], differs all on 2.7 % from radius of an exterior core of the Earth, equal 3485.7 km. If imagine, that the internal structure of protomars before a disrupture resembled structure of the Earth, i.e. had a hot core, liquid mantle and bark, but in other proportions. Then at a disrupture of protomars it lost liquid mantle and solid bark. A part of this bark as its asymmetric satellites Phobos and Deymos we observed now. And satellites are rotated in an equatorial plane of Mars, that shows their birth from a subsoil of protomars. It is possible, that before the Mars disrupture there were conditions for life existence. Because Mars is older than the Earth on 1-2 billion years, the forms of this life should be rather high. After a disrupture of protomars there was a probability that the pieces of this life as DNA came to the Earth, Venus and some satellites of exterior planets. May be only on the Earth the conditions of further development marsian life were favourable.

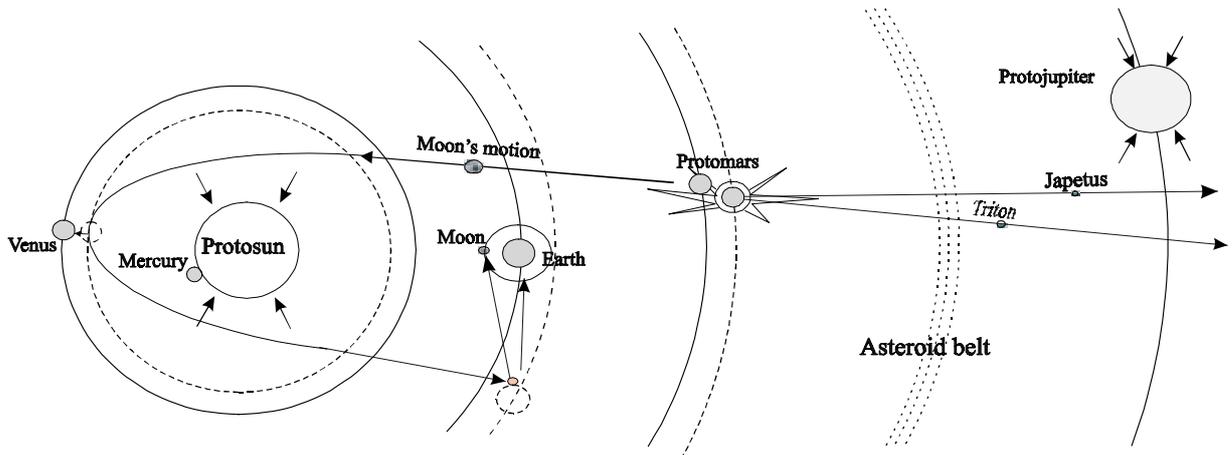

Let us consider the events development after disrupture of the protomars. The part of protomars with mass of 3.766 $M_{Moon}$ began to drop on the sun (see Fig.). This part at the initial moment should have speed, sufficient for a escape from a gravitational field of protomars. The escape velocity of protomars in that time was 6.3 km/sec. Flying up to an orbit of Venus, this part had velocity 11.97 km/sec, equal difference of orbital velocities of Venus and Mars. Near Venus, this fraction had obtained velocity, equal 18.27 km/sec relatively to Venus. There was a contact impact between Venus and this body. But as its velocity was more than escape velocity on a surface of Venus (10.25 km/sec), this body, recoiled from Venus, could turn in the opposite direction – to the Earth. At such contact the direction of Venus rotation around its axis could vary to opposite one, what we observed now. Besides the orbital plane of Venus also was displaced relatively to equatorial plane of the sun on 3,62°. In this case there was an off-centre inelastic impact of two bodies. As follows from the table of orbital velocities, Venus had been given adding energy, and the major semiaxis of its orbit increased on 5.16 million km.

The rejected body began to move in the direction of the Earth with an initial radial speed no more than 8 km/sec. It is complicated to define this value precisely, so it is necessary to calculate a coefficient of restitution, i.e. to find a share of energy, which transformed in in thermal energy and rotational one around Venus axis. Flying up to the Earth, the radial speed of a body was 0.84 km/sec or less, and essentially body came to the same orbit, on which one the Earth moved. Further the body was entrapped by gravitational forces, and began to drop on the Earth with the second solar escape velocity for an earth surface. 10.93 km/sec. After impact with the Earth, the body together with it displaced on more low-altitude orbit - on 11.4 million km. Thus the difference between theoretical and experimental orbital velocities of the Earth was 1.05 km/sec. At the impact to the Earth a part of the body broken away and became a Earth satellite – the Moon with average orbital velocity of 1,05 km/sec. At the impact to the Earth velocity of its rotation around axis considerably increased. And axis has turned on 23° relatively to orbital plane. Orbital plane also displaced from an equatorial plane of the sun on 7°. In addition the declination of a lunar orbit to an ecliptic makes 5,15°, and orbit of Mars 1,85°, the sum of angles is 7°. This magnitude coincides with inclination angle of the Earth orbit to plane of solar equator.

As now average orbital velocity of the Moon is1.023 km/sec, it is possible to consider, that Moon had kept away from the Earth on 5.3 % - 20.5 thousand km for 4 billion years or 5,1 mm per one year. In this case energy of the Earth is transmitted to Moon. This fact is observed experimentally on the Earth as an exceeds of humps of tidal bores in ocean. This wave exceeds motion of axis linked centres of the Earth and the Moon on 2.16° [5].

Other part of protomars moved in the direction of Jupiter. It was represented group of liquid and solid pieces. Large planets entrapped the small-sized parts. They can be observed by us as satellites of planets on high-altitude orbits, which one are move close to plane of ecliptic. Two large pieces were entrapped by Saturn and Neptune is Japetus and Triton accordingly. As the declination of the Mars orbit relatively to plane of ecliptic is 1.85° and Neptune one is 1.77°, the probability of Triton capture by Neptune is rather high. At possible impact of a Triton on Neptune a chunk from Neptune has presumably broken away. It had left on a self-maintained orbit. The plane of this orbit is considerably (on 17°) deflects from plane of ecliptic. We know it as Pluto. Triton began rotate in the counter party relatively gyration of Neptune. Thus the orbits of Neptune and Pluto should be intersected in a point of collision, that we observe.

## 3. External planets of the solar system.

Let us remark, that the sum of masses of Saturn, Uranus and Neptune make 127,0 masses of the Earth. Besides according to the table major displacement of Saturn took place, and rather small aberrations of orbits of Uranus and Neptune from quantum one had taken place. It is possible to consider that earlier protosaturn represented a planet by a mass in 127 $M_{Earth}$. It composed of modern Uranus and Neptune. Last one represented more dense core of the Protosaturn (densities of Saturn equals 0,69 g/cm$^3$, Uranus equals 1,64 g/cm$^3$, Neptune equals 1,26 g/cm$^3$). Probably about 4 billion years ago during sharp transition of the prpotojupiter and the protosun in a new quantum state (presumptively diminution of the sizes of the protosun from the sizes of an orbit of Venus down to an orbit of a Mercury) the strongest gravitation perturbations beat out a core (Uranus and Neptune) from protosaturn. Thus the Saturn came near to the sun on 221,6 million km.

Let us suppose that the energy lost by protosaturn was spent to displace Uranus and Neptune to their modern orbits. According to calculations, only this energy does not sufficient for displacement of Uranus and Neptune. We consider that the necessary part of energy was taken from energy of planet rotation, which coincides on an order of magnitude total kinetic energy of Saturn motion.

Presuming that the planets have a cavity inside, according to paper [2] the squeezing of a planet is determined by following expression:

$$\sigma = \frac{\omega^2 R^3}{2MG}\left[1 - \frac{3}{5}\frac{1-(r_1/R)^5}{1-(r_1/R)^3}\right]^{-1} \qquad (9)$$

where $\omega$ is angular velocity of a planet, $M$ is its mass, $r_1$ is hollow radius, $R$ is Substitute experimentally obtained data of planets squeezing from [3] in the Eq. (9), and we discover the roots of a polynomial of the fifth degree rather $r_1/R$ in limits of [0,1]. For the Saturn and Neptune $\frac{r_1}{R} = 0$, and for Uranus equals 0,72. Thus as a first approximation for Saturn and Neptune the matter focused closer to centre of a planet, and Uranus has cavity of radius not less than 18300 km. It is possible, that Saturn and Neptune has small central cavities similar to an internal core of the Earth [1].

According to [5] Saturn and Neptune is radiated more energy, than immerse from the sun. And for Uranus the balance of immersed and radiated energy is observed, and its effective temperature equals 55K. It testifies to existence of internal power source for Saturn and Neptune and absence those for Uranus.

Uranus has major enough cavity inside. There are the electric currents creating a magnetic field in the cavity. Probably, the currents flow not on equator of the cavity, but are shifted. Thereof its magnetic axis also considerably shifted relatively to the centre of the Uranus. By analogy with paper [1] temperature inside Uranus can be calculated as motion of a current of electrons relatively to Protons.

$$T = \frac{m_e v_1^2}{2k} = \frac{m_e (\alpha^2 c)^2}{2k} = 8.408 K \qquad (10)$$

where $k$ is Boatsman constant, $m_e$ is electron mass, $v_1$ is first solar escape velocity on a surface of Mars, which experimental value is equal to 5.6 km/sec. It is necessary to mark that $\alpha^2 c$ = 15.964 km/sec and it practically coincides the first solar escape velocity on a surface of Uranus. From here follows, that interior temperature of Uranus gives the minor contribution in comparison to energy immersed from the sun, which equals 47K. Analogously it is possible to calculate temperature of other external planets, but it is necessary to take into account, that the measurings of effective temperature are carried out with a major error because of a cloudy coverage of these planets.

There is a problem: whether not protosaturn similarly to the protosun could to pass in a following quantum state, left Uranus, which "had thrown out" the Neptune? Whether present by itself Saturn, Uranus and Neptune a time sweep of a damping star protosaturn?

**4. Conclusions**

The consequences of the new scientific approach introduced in paper [1] with reference to planets the solar system are explored. For internal planets the hypothesis explaining changes of their orbits and redistribution of their masses as of quantum numbers is represented. Is shown, that the consequences of the theory are the possibility of the protomars disrupture under an acting of gravitation forces of the protosun and the protoJupiter at their united squeezing. Thus the part of protomars moving in the direction of the sun, was entrapped by the Earth (including the Moon), other part has formed an asteroid belt and some satellites of exterior planets of the solar system. Is shown, that the part of protomars was entrapped by Neptune as the satellite called as Triton, and at their impact Pluto was formed.

The ideas to a problem on a origin of Uranus and Neptune, as constituents damping star – Protosaturn are expressed. Is shown, that from all exterior planets only Uranus has major cavity inside. As a corollary it there is no interior power source, and the considerable bias of a magnetic axis concerning a rotation axis is observed.

In further it is necessary to improve calculation of redistribution of orbits of external planets at the expense of interchanging between them by momentum and masses. It will give the possibility to track process of fading of stars more precisely.